\documentclass[runningheads]{llncs}

\makeatletter
\newcommand{\@chapapp}{\relax}%
\makeatother

\usepackage{graphicx}
\usepackage[hyphens]{url}
\usepackage{placeins}
\usepackage{algorithm}
\usepackage[noend]{algpseudocode}
\usepackage{graphicx}
\usepackage{booktabs,chemformula}
\usepackage{amsmath,amsfonts,amssymb,epsfig,epstopdf,url,array}
\usepackage[toc,page]{appendix}

\begin{document}
\title{MaaSim: A Liveability Simulation for Improving the Quality of Life in Cities}
\titlerunning{A Liveability Simulation for Improving the Quality of Life in Cities}
%
\author{Dominika Woszczyk \inst{1}
\and Gerasimos Spanakis\inst{1} 
}%

%
\institute{Department of Data Science and Knowledge Engineering\\
Maastricht University\\
\email{d.woszczyk@student.maastrichtuniversity.nl\\ jerry.spanakis@maastrichtuniversity.nl}\\}
\maketitle              

\begin{abstract}
Urbanism is no longer planned on paper thanks to powerful models and 3D simulation platforms. However, current work is not open to the public and lacks an optimisation agent that could help in decision making. 
This paper describes the creation of an open-source simulation based on an existing Dutch liveability score with a built-in AI module. Features are selected using feature engineering and Random Forests. Then, a modified scoring function is built based on the former liveability classes. The score is predicted using Random Forest for regression and achieved a recall of 0.83 with 10-fold cross-validation. Afterwards, Exploratory Factor Analysis is applied to select the actions present in the model. The resulting indicators are divided into 5 groups, and 12 actions are generated. The performance of four optimisation algorithms is compared, namely NSGA-II, PAES, SPEA2 and $\epsilon$-MOEA, on three established criteria of quality: cardinality, the spread of the solutions, spacing, and the resulting score and number of turns. Although all four algorithms show different strengths, $\epsilon-MOEA$ is selected to be the most suitable for this problem. 
Ultimately, the simulation incorporates the model and the selected AI module in a GUI written in the Kivy framework for Python. Tests performed on users show positive responses and encourage further initiatives towards joining technology and public applications.

\end{abstract}

\section{Introduction}
Liveability, wellbeing, quality of life: Those are the concepts that  have been at the centre of a growing interest of industries and governments for the past years. As a matter of fact, multiple scores and rankings have been created, in an attempt to capture the features that describe a ``good quality of life" \cite{Toronto,MUTOPIA,Leefb}.
The focus is nowadays on building and improving cities, and creating the best environment to live in, but also on identifying critical zones that demand changes\cite{Toronto}. Liveability serves now as an evaluation metric for policies.

Furthermore, with the advances in technology, it is now possible to visualise the impact of policies through models and simulations. It allows for cheap analysis, fast insights and no physical consequences. Additionally, stakeholders are given a platform where it is possible to represent concrete plans, thus facilitating the communication and exchange of ideas.   
What is more, it is not rare to see civilians willing to take matters into their hands and take care of projects for their neighbourhood.\footnote{\url{http://www.emma.nl/sites/www.emma.nl/files/tno_community_policing_in_the_hague.pdf}}
Nevertheless, simulators and models for urbanisation are built by companies or collaborating universities. Those are intended for the private sector or a subscription fee must be paid. Moreover, the companies do not share their model, and a user-friendly interface is not their primary concern.
On the other hand, serious games for urban planning, while appealing and entertaining, do not carry a real value regarding practical insights.

Finding sets of beneficial actions that will improve a liveability score is an optimisation task. Hence, a step further for these simulations would be to introduce a decision maker or a decision helper. Real life applications often embody multiple parameters that must be optimised, yet often contradict each other. Implementing an AI module for that simulation requires an algorithm capable of solving multi-objective optimisation problems. Advances in Computer Science brought evolutionary algorithms, widely and successfully applied to optimisation problems and have been shown to be efficient in solving those with a higher number of functions \cite{gamulti}. 

This paper aims to build a simulation model based on a real liveability score and open-source geographical data. Moreover, it proposes an optimisation algorithm that computes an optimal set of actions for achieving best possible liveability score, for a given neighbourhood or municipality. Finally, it combines the model, the AI module and a graphical interface into a serious game, targeted at citizens and policy-makers.     
This paper aims to join those points together and apply it to the new demand for urbanisation and wellbeing, made available to the public. It is a report of the steps and literature for the creation of a serious game, based on real data of Dutch neighbourhoods, from the region of Limburg.

Our contribution is two-fold: (a) we show how a minimalistic city-builder-like simulation can be built based on real data and implemented as a serious game (MaaSim) available to policy makers and (b) how AI algorithms can be used to find the optimal actions to improve a neighbourhood within the simulation framework.

\section{Related Work}
\subsection{Urban Planners and Serious Games}
Urban planners and policy simulators are already present in multiple forms in the private sector. They share a common ground of presenting a 3D visualisation of a city or a region and the effect of performing actions, on different indicators, whether it is liveability, traffic, density, and so on. However, they do not provide a decision maker and are paid services. Some of the current project are:  \textit{Urbansim}\footnote{\url{http://www.urbansim.com}}, \textit{MUtopia} \cite{MUTOPIA},\textit{Tygron}\footnote{\url{http://www.tygron.com/en}} or \textit{SimSmartMobility}\footnote{\url{https://www.simsmartmobility.nl/en/}}.

A serious game is a game with an informative or educational purpose. It can be built on a high level or be very complex. Moreover, the platform is not limited to computer-based programs but can as well be in the form of board games. In this case, serious games aim at urban project stakeholders or civilians. The goal is to encourage a thought process and teach about some neglected or taken for granted aspects of urban planning. 

Some current games built for the purpose of urban planning were made by means of 3D simulations, like the ``B3: Design your marketplace''project \cite{B3} or games made by the Tygron company \cite{tygron}. Others were made in the format of a board game \cite{sterdam} or card games. \footnote{\url{https://www.kickstarter.com/projects/ecotopia}} 
Nevertheless, those games are targeted for universities and business applications. Moreover, the access is restricted to specific events and test groups \cite{tygron}.   
  
More simplistic serious games not based on real models aim to educate about the difficulties of managing the different aspects of livability and environment, as well as raise awareness.
\footnote{\url{https://www.conservation.org/NewsRoom/pressreleases/Pages
/Ecotopia\_Launch\_Announcement.aspx}} 
\footnote{\url{http://electrocity.co.nz/Game/game.aspx}} \footnote{\url{https://techcrunch.com/2011/06/08/multiplayer-facebook-game-trash-tycoon-trains-you-to-be-green-but-in-a-fun-way} }

\subsection{Paper Application and Structure}
This paper aims at filling the gap between serious games and urban planners by presenting a report on the construction of a 2D liveability simulation. 

First of all, the creation of the simulation follows scientific methods and real-life dataset and score, as described in Chapters 3 and 4. On that aspect, the paper situates itself among other urban planners mentioned in the previous section. On the other hand, the open source essence of the final product and the raised attention to a user-friendly and entertaining interface places it among other serious games. Finally, the addition of an optimisation algorithm (described in Chapter 5) as a decision helper is an unusual asset that is present in none of the urban planners or serious games. This AI module shows the user the best possible choices by giving sets of optimal solutions. In that manner, the user can select their favourite solution based on its criteria. They are offered a choice instead of having one imposed on them.  

The end product of this paper will benefit policymakers by giving them general insights on the development of neighbourhoods, for Dutch and non-Dutch citizens. Most importantly, the end program can bring awareness to the inhabitants and provide a visualisation platform that shows what can be improved, through a playful, serious game. In that manner, civilians can undertake action and propose projects to municipalities. The paper is an invitation to bring data-based visualisation and decision making tools to the public.


\section{The Dataset}

The dataset was based on an existing liveability score for Dutch neighbourhoods. The Leefbaarometer 2.0 is a score built by Rigo, a Dutch statistical company and Atlas, on the demand of the Ministry of Housing, Spatial Planning and the Environment \cite{Leefb}. It is a low-level score based on 116 environmental variables grouped into 5 categories. Those variables were collected based on data given by municipalities and surveys, corrected for different factors such as age and general background. Since the variable values were not available for the public, a different approach had to be taken: Instead of using the exact indicators and scoring function, all available indicators would be collected and computed using the same method, which is using the 200 m area around the centre and distances to facilities. The new scoring function would be then computed using the liveability classes made publicly available by Rigo. Therefore, the following dataset was built based on open data for the whole Limburg, retrieved from CBS\footnote{\label{cbs} \url{https://www.cbs.nl/}}, BAG\footnote{\url{http://geoplaza.vu.nl/data/dataset/bag}}, Geofabrik\footnote{\url{http://download.geofabrik.de}} and Culturelerfolg\footnote{\url{https://cultureelerfgoed.nl}}. Fortunately, multiple original indicators could be discarded and ignored due to the fact that they remain constant for that region. Unfortunately, some were not available, i.e. average distance to an ATM or crimes/nuisance records. 

The indicators were retrieved and computed following the indicators description \cite{Leefb}, using shapefiles. A shapefile is a file format combining information with its geometry(Point, Polygon, Line) and its coordinates \cite{esri}. In this case, the shapefiles contained: buildings and their function (habitation, industry ) and construction date, neighbourhoods, roads, tracks, waters and land use. This format allows to make geographic computations such as distance, centres and areas, as needed for computing the indicators. Moreover, shapefiles allow for easy visualisation and integration within a simulation, needed for the further step of the paper. An adjustment had to be made in the manner the indicators were computed. For many calculations, the centre of the neighbourhood is used. Rigo did not state how it was computed. Therefore, both the geometrical centre and the mean coordinates of all buildings were compared. The mean coordinates method was retained as it was better at reflecting the ``centre" of a neighbourhood. The geometrical variant was often in a non-occupied area.   

The resulting generated dataset consisted of 997 neighbourhoods and 54 indicators.

\section{Model Selection}

The simulation was built on real-life indicators to create in-game indexes, showing the user the state of the selected neighbourhood and its liveability. Furthermore, actions affecting those had to be generated, in a relatively accurate manner. Finally, a liveability score based on the indicators had to be created.   

As not all indicators retrieved were initially significant, the data had to go through a process of selection and combination of features. This section describes the process of the analysis, to reach the final indicators and actions suitable for the simulation and the creation of a liveability score.  


\subsection{Techniques}
\subsubsection{Dimensionality Reduction}
The aim of this section is to simplify the model by creating understandable actions and indicators. For the purpose of grouping and deriving actions, dimensionality reduction techniques are used. Feature selection can be achieved by different methods, depending on the desired output. Several dimensionality reduction and feature selection are available \cite{dimsurvey}.

However, Exploratory Factor Analysis(EFA) \cite{EFA} was more suited  to the problem, thus was retained. EFA is a technique that finds the underlying latent factors that cannot be measured by a single variable and that causes the changes in the observed variables. The model can be interpreted as a set of regression equations

\subsubsection{Indicators and Tests}

Multiple indicators can be used as support for selecting features and improving the quality of the dataset. The following paragraphs describe the tools used and tests performed in the next section.

The Correlation matrix is used to show the pairwise correlation value for variables in a dataset. It can be used to discard variables in the cases of redundancy, when the correlation is too high, and/or when the correlation is too low. It can also be used as a tool to combine variables.

Communalities show the extent to which a variable correlates with all other variables. They indicate the common variance shared by factors with given feature. Items with low communalities can be discarded to improve the analysis.

The  Kaiser-Meyer-Olkin (KMO) is a measure of sampling adequacy. It indicates the proportion of variance in the variables that might be caused by underlying factors. Thus, it can measure how suitable the dataset is for Feature Analysis.

Feature engineering is also necessary for combining variables, re-expressing them into boolean or more explicit indicators.

\subsection{Methodology}
The developed approach is described in detail with the following subsections. 
 
\subsubsection{Feature Selection and Score Creation} 
The scoring function used by Rigo no longer applied to the collected indicators. Therefore, a new score function had to be found. Nevertheless, the liveability classification was available. Thus, a regression could be applied. 

The dataset was firstly examined for constant indicators and outliers. One variable remained constant and was consequently removed. The neighbourhoods have been classified in 6 ordinal classes, the first class having the worst liveability score. 

\begin{figure}[htb]
\includegraphics[width=8cm]{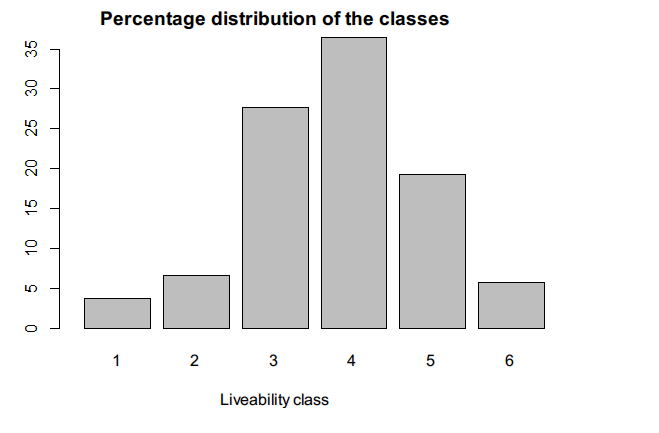}
\centering
\caption{Liveability classes distribution of the dataset}
\label{fig:1}
\end{figure}

As one can see in Figure \ref{fig:1}, the class with the lower score had in total 3 data points. It was decided to discard that class as there were too few data points to guarantee a good prediction.    

In the next step, variables with a correlation of $\mid$1$\mid$ were discarded, based on the indicator description and correlation with other variables.

In Figure \ref{fig:1}, one can observe that the dataset was not evenly spread among classes. The imbalance in a dataset can negatively impact the results of a prediction. Therefore, the Synthetic Minority Over-sampling Technique (SMOTE) introduced in 2002 \cite{SMOTE}, was performed to correct the imbalances by over-sampling the minority classes.
 
In order to find a model for the score, Regression was performed for classification. The algorithm predicted the numerical value of a data-point class(1 to 6), and the actual predicted class was obtained by truncating that number to an integer. This technique had been used in order to later have a scoring function that output numerical values instead of classes, but in order to validate the model, the classification recall was necessary. Different Regression algorithms were compared, namely multinomial regression, K-nearest neighbours(KNN), Random Forest(RF) and a Decision Tree as well as a perceptron.     
In Table 1, one can see that SMOTE improved the recall of all algorithms. The Random Forest(RF) one achieved the best recall, and thus was retained.  
 
\begin{table}
\includegraphics[width=8.5cm]{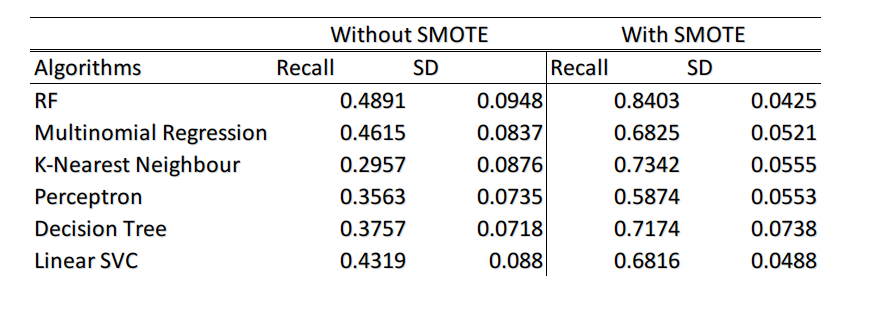}
\centering
 \caption{Multinomial classification algorithms comparison based on a 10-fold cross validation}
\end{table}

 Additionally, to reduce over-fitting, the generated RF was used for feature selection using mean decrease impurity as a feature importance measure. This measure indicates how much each indicator decreases the impurity of a tree, the variance in the case of a regression. The value is computed for each tree and for each feature. For the RF, the importance of a feature is the average of those values \cite{rffeatsel}. Indicators with too little impact were discarded. 
 
 \begin{figure}[ht]
\includegraphics[width=8cm]{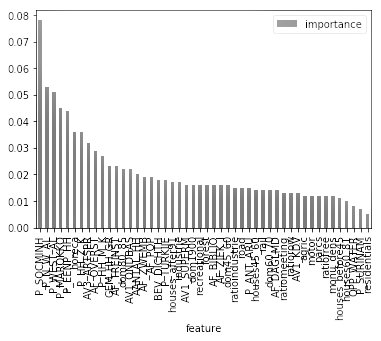}
\centering
\caption{Importance of the features as given by the RF }
\end{figure} 
 
Based on Figure 3, it was decided to discard the last three variables, as the importance dropped noticeably at the last three elements, and their importance value is below 0.01. Once the variables were cleaned, a 10-fold cross-validation resulted in an average recall of 0.83. To interpret the resulting RF, one can decompose the Decision Trees into a prediction function. The function can be described as the sum of the contribution of each feature and the value of the root node. 

The value for a decision tree is computed as follows:

\begin{equation}
f(x)= bias + \sum\limits_{k=1}^{K}contrib(x,k) 
\end{equation}
where K is the number of features and bias is the value at the root of the tree.

For a Random Forest, the prediction value is the mean of all the values of all Decision Trees \cite{tree}.

\begin{equation}
F(x) = \frac{1}{J} \sum\limits_{j=1}^{J} bias_j + \sum\limits_{k=1}^{K}(\frac{1}{J}\sum\limits_{j=1}^{J}contrib(x,k)) 
\end{equation}

The model was saved and used as an evaluation function. The number of indicators after the process was 44.

\subsubsection{Actions and Groups}
Exploratory Factor Analysis(EFA) was performed in order to find actions affecting the indicators, as they can be described as underlying variables that affect the observable variables. After dropping variables with low communalities, the KMO value of the dataset was 0.7, which was acceptable. 
The indicators were divided into two groups: direct and indirect. The indirect indicators were ones that cannot be directly controlled by an action. For example, the percentage of single households, non-westerners and family households. The directs indicators were the ones that can be directly ``added"  such as groceries stores, parks, swimming pools, and others. An action ``Add a library" increases or decreases the indicator by a pre-computed value. For each ``Add an item" action, the value increase by 10\%  of its original value, and similarly for a decreasing action.
Another way to compute the effect of an action has been investigated, namely computing the increase by adding one element of that specific type at the density centre of the neighbourhood. However, as the density centre agglomerates most of the inhabitations, the increase on the liveability score were too small to be considered. An increase/decrease of 10\% was big enough to show significant changes but small enough not to compromise the feasibility of the actions. It was decided on keeping 1 indirect action and 11 direct ones.

Finally, to reduce the number of indicators shown in the simulation, variables were grouped into categories. Through the process, 5 groups were formed. These grouping's only purpose was to alleviate the interface. Consequently, the author took the liberty to select the indicators representing similar features, i.e. services, environment, housing, healthcare and leisure. Nevertheless, the newly formed indicators had to be meaningful to the user. Therefore, the normalised contributions of each feature, as computed by the RF model and described in equation(2), were used to weight the indicators.       


\section{Optimization Algorithms}
This section presents and describes multi-objective optimisation algorithms(MOOA). Four algorithms were compared on several metrics, and the one judged the most adequate was implemented into the simulation.  

\subsection{Problem Definition}

The optimisation problem of this paper can be defined as:

\begin{equation}
\begin{aligned}
&{\text{Maximise}}
& & f(x)= s \\
&{\text{Minimise}}
& & f(y)=t \\
& \text{subject to}
& & t > 0
\end{aligned}
\end{equation}

where $s$ is the liveability score and $t$ the number of turns (or actions).

The problem stated above is a Multi-Objective Optimisation problem (MOOP) and'as opposed to a single optimisation problem, contains more than one variable that needs to be optimised. It can be homogeneous if all variables need to be maximised/ minimised, or mixed when it is constituted of both minimisation and maximisation functions, also called minmax \cite{minmax}. The above-stated problem both minimises and maximises its variables, thus it is mixed.  

The problem was translated into a string of binaries for each possible action and for each neighbourhood. A value of 1 indicating the corresponding action is used, and the total number of 1's indicates the number of turns. Then, each indicator affected by an activated action is updated. Finally, the new liveability score of an individual is computed using the model described in 4.2.  






\subsection{Algorithms}
Multi-objective optimisation algorithms have been a popular research topic, with a growing interest. There are plenty of research papers covering those algorithms compared and tailored for their specific application problems \cite{pindoriya2010comprehensive,hydro}. Hundreds of variations have been proposed, some specific to a particular type of application. They can be pareto-dominated based, indicators based, probability based, preference-based or swarm based \cite{pindoriya2010comprehensive}. A comprehensive survey can be found in \cite{qu2018survey}. As described in 5.1, the problem of this paper is discrete.  Four well-known optimisation algorithms for static and discrete multi-optimization problems were compared and are briefly described here.

\paragraph{ Non-Dominated Sorting Genetic Algorithm (NSGAII)}  Published in 2002 by  Deb et al. \cite{NSGAII}, an improved version of the original NSGA by Guria \cite{guria}, it is a non-dominated Pareto optimisation algorithm. It is the most used and well known algorithm due its to simplicity, diversity-preserving mechanism, and better convergence near the true Pareto optimal set. 

\paragraph{ Pareto Archived Evolutionary (PAES)}Published in 2000 by Knowles and Corne \cite{paes}. It is a simpler method, as it does not have a population but keeps non dominated solutions in an external archive. At each iteration, a parent solution is mutated, and a solution $c$ is created. The parent $p$ is then evaluated and compared to the mutated solution $c$. If $c$ dominates $p$, then $c$ becomes the new parent and is added to the external archive.

\paragraph{Strength Pareto Evolutionary Algorithm(SPEA2)}
Zitzler et al. \cite{spea}, introduced in 2001 as an improved version built upon the SPEA, proposed earlier by the same group Zitzler and Thiele \cite{spaeone}. SPEA2, also uses an external archive, but uses a truncate procedure that eliminates non dominated solution but preserves the best solutions, when the size of the archive is exceeded.

\paragraph{$\epsilon$-Multi-objective Evolutionary Algorithm ($\epsilon$-MOEA)}
$\epsilon$-MOEA presented in 2003 by Dieb et al. \cite{moea} is a steady-state algorithm, that is it updates the population one solution at a time and coevolves a population of individuals with an external archive, where are stored the best non dominated solutions. The external archive applies $\epsilon$-dominance to prevent the deterioration of the population and to ensure both convergence and diversity of the solutions. 

\subsection{Metrics}
Genetic algorithms for multi-objective optimisation problems(MOOP) cannot be easily compared based on traditional measures used for single-objective problems. The quality of a set of solution given by a multi-objective optimisation algorithm is defined by the distance from the true Pareto Front(PF), its coverage and the diversity of the solutions. Therefore, well-known measures for MOOP were used \cite{metrics}.

\paragraph{Spread} The spread, also called ``extent", is a diversity and coverage measure that shows the range of values covered by the set of solutions. \cite{hamdy2016performance}    
\begin{equation}
 \resizebox{100pt}{!}{
$\Delta = \frac{d_f +d_l + \sum\limits_{t=1}^{N-1} \mid d_i - \bar{d_l} \mid}{d_f + d_l + (N-1)\bar{d}}$}
\end{equation}

where $d_i$ is the Euclidean distance between two consecutive solutions in $PF$ and $\bar{d}$ is the average of these distances. For a perfect distribution, $\Delta$ =0 which means that $d_i$ is constant for all i.

\paragraph{Spacing} The Spacing indicator is a distance-based measure of the uniformity of a solution. \cite{hamdy2016performance}
For a set of solutions S the spacing is defined as

\begin{equation}
 \resizebox{120pt}{!}{
$SP(S) =\sqrt{\frac{1}{\mid S-1 \mid } \sum \limits_{i=1}^ {\mid S \mid} (\bar{d}-d_i)^2}$}
\end{equation}

where $\bar{d}$ is the average of $d_i$ and $d_i$ is the Euclidean distance between and solution s and the nearest member in the true Pareto Front.

A value of 0 indicates that the solutions are evenly spread along the Parieto Front. 

\paragraph{Cardinality} is a measure of the number of non dominated solution given by an algorithm \cite{hamdy2016performance} It is especially relevant to the problem of this paper, as an algorithm offering multiple scenarios is preferred. 

Additionally, the maximum score and the minimal number of turns from a set of solutions are  measured.

\subsection{Experiments}
\subsubsection{System Specification}
The Platypus library for Python\footnote{\url{https://github.com/Project-Platypus/Platypus}} was used for the implementation of the optimisation algorithms. All experiments were performed on an i7 7700HQ Intel Core Processor 2.80GhZ and in Python 3.6.

\subsubsection{Experiment Set Up}

\begin{table}[h]
\centering
 \resizebox{190pt}{!}{
  \begin{tabular}{@{}cc@{}}
    \toprule
    Algorithm & Parameters \\
    \midrule
    \ch{NSGAII} & pop size = 100\\
    \ch{PAES} & pop size = 100, archive size=100\\
    \ch{SPEA 2} & pop size = 100, archive size=100\\
    \ch{$\epsilon$-MOEA} & pop size =100, archive size=100, $\epsilon$ = $0.01$\\
    \bottomrule
  \end{tabular}}
  \caption{Parameters setup}
\end{table}

The algorithms were compared on 10000 and 20000 function evaluations(FE). Below 10000, the results were not as interesting in terms of fitness and competition among algorithms, while more than 20000 would necessitate greater computational power.

\subsubsection{Results}
\begin{figure}[htb]
\includegraphics[width=14cm]{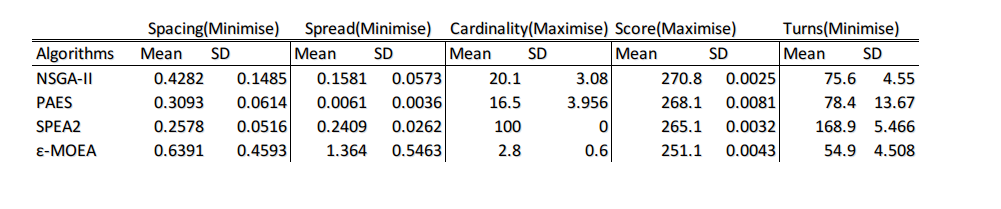}
\centering
\caption{Results on 10 runs for 10000 FE}
\end{figure} 

\begin{figure}[htb]
\includegraphics[width=13cm]{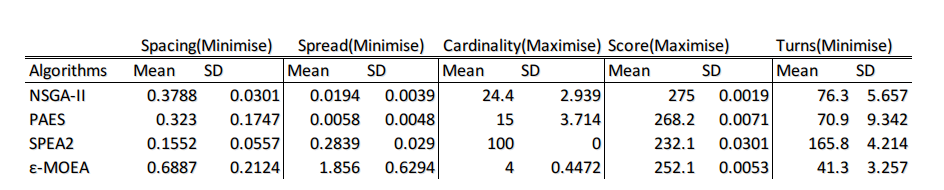}
\centering
\caption{Results on 10 runs for 20000 FE}
\end{figure} 

The Wilcoxon-Mann-Whitney's test was performed on the 20000 FE runs to check whether an algorithm is better than another one \cite{wilcoxon}. The test was performed for each metric. A cross is present whenever an algorithm in the row was significantly better than the algorithm in the column.

\begin{figure}[htb]
\includegraphics[width=12cm]{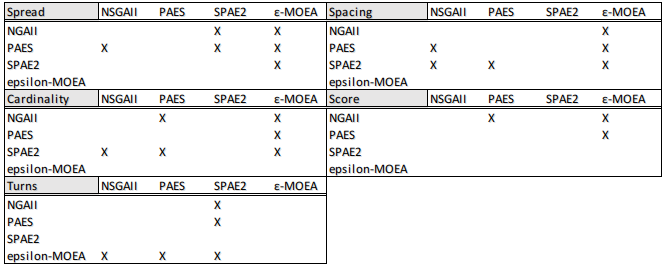}
\centering
\caption{Results of the Wilcoxon-Mann-Whitney's test }
\end{figure} 



\subsection{Discussion}
One can notice that the cardinality of the SPEA2 algorithm was always equal to 100. This can be explained by the fact that the external archive limits the total number of solutions SPEA2 retain. If the archive size was increased, it could results in an even higher cardinality. Nevertheless, it is surprisingly high compared to other algorithms, considering that only unique non dominated solutions were counted.

One can observe on Figure 6 $\epsilon$-MOEA algorithm was outperformed in almost all metrics and by nearly all the other algorithms. Indeed, $\epsilon$- MOEA  favoured faster convergence, at the expense of diversity and uniform spread. However, considering that it reaches a fairly similar score in a significantly smaller number of turns, it wouldn't be correct to classify it as ``poorly performing".

Both NSGA-II and PAES performed well for different measures, sometimes even outperforming each other. While PAES had better results for the spread and spacing, NSGAII gave higher scores in fewer turns.

On the other hand, SPEA2 was the best out of the four algorithms when comparing spacing and cardinality. However, the number of turns needed was significantly higher, while achieving a similar score as the other algorithms in Figure 4, and a lower one in Figure 5.  

Ultimately, a relatively high score in fewer turns was in the interest of this paper application, more than diversity and uniform coverage of the Parieto Front. Particularly when each turn represents a lot of money and work.   
Therefore, the decision of choosing the  best optimization algorithm was between the NSGA-II and $\epsilon$-MOEA.

Finally, it was decided that $\epsilon$-MOEA would be most suited for this paper problem. The characteristic faster convergence of the algorithm fits best the simulation environment, as users do not like to wait.

\section{Simulation}
The final product, the simulation, is the integration of the previous sections. The scientific methods described earlier are represented in darker shades.

\begin{figure}[ht]
\includegraphics[width=9cm]{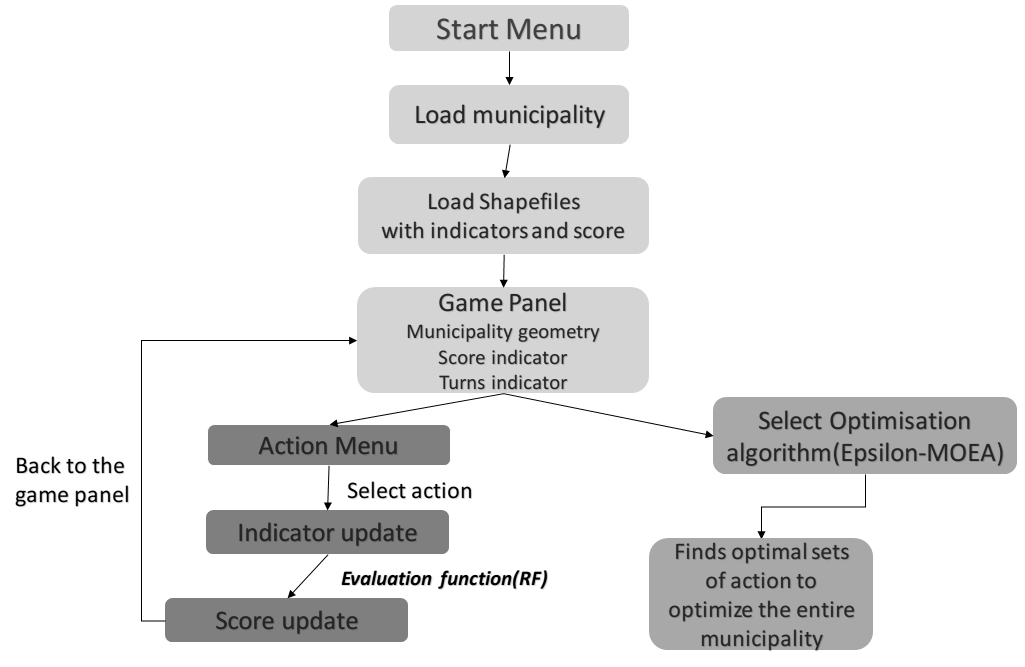}
\centering
\caption{Pipeline of the simulation }
\end{figure}

The interface (GUI) was implemented in a minimalist and intuitive way, in Kivy for Python. 
An AI option has been added to the simulation menu. Once started, the algorithm outputs the set of solution of optimal actions. The dataset and code for the application will be made available upon paper acceptance.

\subsection{Experiments}
In order to assess the effect of the simulation, experiments with users were performed in an interview fashion. Users were asked to play around with the simulation. Then, they were assigned a task of reaching the highest score possible within 10 turns. After the test, the user had to grade the value of the game in term of entertainment, education, aesthetics and the ergonomics of the controls. The grade given was in a scale from 1 to 10, 10 being the highest grade.   

\subsection{Results}

\begin{table}[h]
  \centering
  \resizebox{200 pt }{!}{  
  \begin{tabular}{@{}cccccc@{}}
    \toprule &Entertainment & Education & Aesthetics& Ergonomy& Score \\
    \midrule
    \ch{Mean} & 7.1 & 7.4 & 8.1 & 6.9 & 36.5\\
    \ch{SD} &  1.28 & 1.17 & 1.19 & 0.8 & 18.5\\ 
    \bottomrule
  \end{tabular}
  }
  \caption{Results from 10 users tests}
\end{table}

From the results on Table 3, one can observe that the mechanics have room for improvement but that overall, the different aspects are positively graded. There were comments regarding the playability, difficulty due to too many choices and the symbols, that were not always understood. Nevertheless, users were overall pleased with the design, the controls and the purpose of the simulation. 

\section{Discussion}
Of course, some limitations and critiques can be addressed concerning the process of creating the simulation. One first possible limitation of the model described in chapter 4 is that it was built on a dataset of the region of Limburg.   The dataset could have been extended to the whole Netherlands. However, due to computational difficulties, a larger dataset was not possible.

The liveability model has undoubtedly room for improvement. For example, more open source data could have been collected and used in the regression model. Resulting actions could perhaps be even more meaningful. Hopefully, the resulting framework can be easily modified to incorporate a new scoring function, for future changes.

Moreover, one can argue that the interface could be improved for an even more appealing serious game of a higher educational value. However, as it was not the primary focus of this paper, the GUI was kept simple.

\section{Conclusion and Future Work}
As liveability becomes a key factor in everyday life and data science techniques advance, it is only natural to combine them in order to improve decisions for the good of society. This paper reports the methodology followed to build a serious game for urban planning and on incorporating an AI module as a decision helper.   

First, the steps followed to create a model for a simulation were described, based on real data and scientific techniques. To create a new score with open-source data, Exploratory Feature Analysis(EFA) and Random Forests(RF) were both used to prune non-significant variables. Then, RF for regression was applied and achieved a recall of 0.83 with 10-fold cross-validation, with a reduction from the original 115 to 44 indicators. The indicators were divided into direct and indirect action categories; the indicators that are directly related to an action and the ones that are the indirect result of an action. The latter were created by looking at underlying factors using EFA. This process resulted in 12 actions, 11 directs and 1 indirect. Moreover, the model was built in a suitable manner for the simulation. For visibility purposes, the indicators were gathered manually into 5 groups: housing, environment, services, healthcare and leisure.  

Later on, the optimisation problem resulting from the model was formulated as a multi-objective optimisation problem. The algorithms NSGAII, SPAE2, PAES and $\epsilon-MOEA$ were compared. The algorithm that was the most suited for the simulation was $\epsilon$MOEA due to its fast convergence and higher results in terms of turns and score. Research works covering high dimensionality multi-objective optimisation problems (many-objective optimisation problems), could be investigated in the future, since dimensionality increase might lead to a more complex optimization problem. Furthermore, state-of-the-art algorithms but more complex such as Binary Bat Multi-Objective Algorithm or Multi Objective Ant Colony Algorithm could be compared to algorithms in this paper. 

Finally, the model and the optimisation algorithm were incorporated into an interface. The feedback received from testers was generally positive. Nevertheless, further improvements in the interface can only increase the quality of the serious game.

The resulting simulation from the process of building the model, the AI module and the interface answered the questions on how to incorporate an AI module for a simulation and how to build a serious game based on real data. In light of these results, one can conclude that data science techniques can be successfully applied to building an urban planner for both entertainment and education purposes. Arguably, the product of this paper is one step forward in filling the gap between private and public applications.

\bibliographystyle{splncs04}
\bibliography{lib.bib}


%
%
%

%




\end{document}